\documentclass{article}
\usepackage{graphicx}
\usepackage{amsmath}
\usepackage{amsfonts}
\usepackage{amssymb}

\begin{document}
\bigskip

\begin{center}
\bigskip Novel Scheme for Universal Quantum Computation

C. M. Bowden, and S. D. Pethel

Weapon Sciences Directorate, AMSAM-RD-WS-ST

Research, Development, and Engineering Center

U. S. Army Aviation and Missile Command,

Redstone Arsenal, AL 35898, U. S. A.

email: cmbowden@ws.redstone.army.mil

Abstract
\end{center}

\begin{quotation}
A scenario for realization of a quantum computer is proposed consisting of
spatially distributed q-bits fabricated in a host structure where nuclear
spin-spin coupling is mediated by laser pulse controlled electron-nuclear
transferred hyperfine (superhyperfine) Fermi contact interaction. Operations
illustrating entanglement, nonlocality, and quantum control logic operations
are presented and discussed. The notion of universality of quantum computation
is introduced and the irreducible conditions are presented. It is demonstrated
that the proposed generic scenario for realization of a quantum computer
fulfills these conditions.

International Journal of Laser Physics, January1, to appear.

\bigskip
\end{quotation}

\section{I. Introduction}

The Nobel Laureate Richard Feynman first suggested that a new type of
computer, running on the principles of quantum rather than classical
mechanics, might have an exponential speed-up in computation time over a
classical device for some tasks[1]. This observation engendered the
development of a theory of a digital quantum computer by David Deutsch of
Cambridge University[2], in which he demonstrated the intrinsic massive
parallelism of quantum computation by operating on a coherent superposition of
a large number of states. In particular, a single computation acting on N
quantum bits can achieve the same effect as 2N simultaneous computations
acting on classical bits. For N=300, this is a number greater than the number
of atoms in the visible universe. Obviously, no classical computer could ever
be built to compete with this kind of processing power. Exact methods and
applications that harness this enormous potential are currently the subject of
intense research. In 1994 Peter Shor of Bell Labs announced the discovery of
an algorithm which, if running on a quantum computer, could find prime factors
of large numbers in polynomial time[3]. There is currently no known classical
counterpart that can accomplish the same task in less than exponential time.
Shor's factoring algorithm was the impetus for an explosion of research aimed
toward the theory and experimental implementation of a working digital quantum
computer[4]. In a subsequent development, Lov Grover discovered an algorithm
that shows a square-root speed-up over a classical machine for the task of
searching a large unordered database[5]. Even though Grover's algorithm is
only polynomially faster than a classical algorithm, the wide range of its
applicability to data search makes it very enticing. Another extremely
important class of problems is represented by the so-called ''intractable''
NP-complete set. These problems are at the core of a wide range of practical
routing, layout, and other logistics issues. A prototypical example here is
the traveling salesman problem (TSP). It is widely believed that no classical
algorithm that can solve this problem, or any other member of this class,
exactly in less than exponential time. Such problems are now a challenge for
quantum computers and algorithm developers.

Various approaches are currently in process for laboratory realization of
essential elements to perform quantum computing, namely, the construction of
qubits and demonstration of quantum logic operations. The work of Jeff Kimble
and his colleagues at the California Institute of Technology (Cal Tech)
emphasizes the use of entangled photon states using microcavities to construct
quantum logic gates [6]. Correspondingly, David Wineland and his group at
NIST, following the theoretical predictions of Peter Zoller and coworkers[7],
achieved what amounts to a two bit quantum register using laser excitation
control of atoms in cold ion traps[8]. These methods require large scale,
state-of-the-art laboratory facilities and defy scalability beyond a few
qubits. Aside from the requirements of extremely sophisticated laboratory
facilities and procedures, quantum decoherence occurs in these schemes at a
time scale of nanoseconds at best, and is a major obstacle. Recently, Neil
Gershenfeld and Isaac Chuang introduced a revolutionary scheme for quantum
computing using nuclear spins and the methods of nuclear magnetic resonance
(NMR) to construct and manipulate quantum logic[9]. Free temporal evolution
takes place virtually without decoherence due to the robust isolation of
nuclear spins caused by screening by the atoms electrons. This version of
quantum computation can be controlled and processed with current
state-of-the-art technology. More recently, Chuang, Gershenfeld, and Kubinec
(CGK) reported, using this scheme, the first ever laboratory demonstration of
a genuine quantum computation [10] by execution of Grover's quantum search
algorithm [5]. This included loading of the registers, unitary execution of
the quantum calculation, and read-out of the results. Their results, though it
utilized only four input registers, constitute a significant step forward
toward the practical realization of a useful quantum computer. One of the
problems in scaling, in this regard, to a practical device is that the nuclear
spins within a molecule that form the computational basis must rely on
chemical shifts to be distinguished from one another. These shifts are usually
very small and consequently the rf pulses that must control the computation
are required to be sufficiently long so as not to couple distinct spins by
resonance overlap. This is usually of the order of microseconds. Thus,
computation is required to be exceedingly slow. A second, and more severe
problem in scaling is that the read-out signal necessarily diminishes
exponentially with the size of the quantum bit register. The current
state-of-the-art limits this size to no more than 15 bits. We recognized early
on [11] that these difficulties could be mitigated by selectively coupling the
nuclear and electronic intrinsic spin angular momenta via hyperfine or
transferred hyperfine interactions [12,13]. Further emphasis has been recently
enunciated in the use of electron-nuclear spin coupling and control for
quantum computation by David DiVincenzo[14]. These comments were stimulated by
the specific seminal suggestions by Kane[15] for nanostructures in Si. The
scheme proposed by Kane surmounts the problems of scalability and individual
nuclear spin addressability that is inherent in the liquid state NMR methods
[9,10]. However, it suffers in principle from the requirement of electron
charge transfer, and the measurement of single electronic charge by
nanostructured gates for signal read-in and read-out. Otherwise, electronic
charge migration is required for mediation of nuclear spin coupling as well as
the shift of individual nuclear spins in and out of resonance with a constant
rf field. Electronic charge and spin migration is controlled by nanofabricated
gates. The requirements of the procedure are beyond current fabrication
technology, but this is likely to change in the reasonably near future.
However, there remain problems connected with charge transfer control for
nuclear spin identification, and single charge measurements for read-in and read-out.

Our purpose here is to present a scheme for quantum computation that is
universal and that does not suffer from the limiting conditions of liquid
state NMR and that may offer a more viable alternative to the proposed scheme
of Kane. We draw upon the favorable decoherence properties of nuclear spin
systems utilized in NMR quantum computing, and the solid state distributed
implanted nuclear spin elements system coupled via electron hyperfine and
transferred hyperfine interactions proposed by Kane. Our scheme involves a
distributed individually addressable system of electron spin/nuclear spin
qubits and state-of-the-art microwave single electron spin resonance for
signal read-in and sensitive optical fluorescence for read-out. Our treatment
here is generic. Specific materials and optimizations are relegated to a
future publication.

In the next section we present and discuss the irreducible requirements for a
universal quantum computer. Our generic scheme is presented in Section III.
Quantum logic, entanglement, nonlocality and quantum information memory are
discussed in Section IV where it is shown that all the conditions of Section
II are met. The last section is used for discussion and comparison with other
proposed methods and schemes, and for concluding remarks.

\bigskip

\section{II. Requirements for Universal Quantum Computer}

\bigskip

Quantum computational (QC) algorithms are of only esoteric significance apart
from the hardware capability to execute them. The resource and technological
investiture necessary to develop such a device would necessitate design
requirements that fulfill the capability to perform any given quantum
computation and accommodate the execution of any algorithm, including initial
state preparation, read-in and read-out. A quantum computer design meeting
these requirements will be termed universal. The design must meet, therefore,
certain irreducible requirements that are both necessary and sufficient for
the execution of any given QC algorithm. Establishing these requirements
necessitates the identification of the set of elemental operations common to
the execution of all quantum computational algorithms that, in a particular
operational sequence, are sufficient for the execution of any given algorithm.
A universal quantum computer must be capable of execution of the elemental operations.

It is conjectured that the identification of universal elemental operations
common to al QC algorithms and sufficient in specific combination are simple
Hadamard transforms, together with a generalized quantum phase shift operation
[16]. The work of Cleve, et.al. [16] is the basis for this conjecture. A
simple Hadamard transform, $H$ , on a qubit $\left|  q\right\rangle $, $q=0,1$
is given by [16]%

\begin{align}
& \left|  0\right\rangle \overset{H}{\rightarrow}\frac{1}{\sqrt{2}}(\left|
0\right\rangle +\left|  1\right\rangle )\label{eq1}\\
& \left|  1\right\rangle \overset{H}{\rightarrow}\frac{1}{\sqrt{2}}(\left|
0\right\rangle -\left|  1\right\rangle )
\end{align}
This operation is exactly equivalent to a uniform single particle beam
splitter[17]., and is a special case of the more general Quantum Fourier
Transform (QFT),
\begin{equation}
\left|  a\right\rangle \overset{F_{2^{m}}}{\rightarrow}\sum_{y=0}^{2^{m}%
-1}e^{2\pi iay/2^{m}}\left|  y\right\rangle
\end{equation}
In terms of a two-bit operation a subsequent phase shift operation can impart
a conditional phase shift to one component of the transformed wave function,
\begin{equation}
\left|  0\right\rangle \left|  1\right\rangle \overset{H}{\rightarrow}\frac
{1}{\sqrt{2}}(\left|  0\right\rangle +\left|  1\right\rangle )\left|
u\right\rangle \overset{c-u}{\rightarrow}\frac{1}{\sqrt{2}}(\left|
0\right\rangle +e^{i\varphi}\left|  1\right\rangle )\left|  u\right\rangle
\end{equation}
Or a conditional transition of the auxilliary qubit as well,
\begin{equation}
\frac{1}{\sqrt{2}}(\left|  0\right\rangle +\left|  1\right\rangle )\left|
u\right\rangle \overset{C-NOT}{\rightarrow}\frac{1}{\sqrt{2}}(\left|
0\right\rangle \left|  u\right\rangle +e^{i\varphi}\left|  1\right\rangle
\left|  v\right\rangle )
\end{equation}
thus forming an entangled state and constituting a controlled NOT-gate
(C-NOT). Ultimately, quantum trajectories are brought together as in a
Mach-Zehnder interferometer to produce quantum interference[17]. The
intermediate conditional operation, phase shift and/or entanglement, requires
auxiliary qubits and a unitary operation[16]. The corresponding network
representation is given in fig. 1.%

\par\noindent Fig. 1. Network representation of quantum Mach-Zender interferometer with
generalized phase shifter.

\bigskip

Thus, the elemental operations are conjectured to be constituted by a simple
QFT, a conditional phase shift and/or entanglement, and ultimately a QFT that
brings together quantum trajectories to produce quantum interference. The
later constitutes the results of a computation, i.e. collapse of the wave
function. These operations are entirely equivalent to the unique
characterization of quantum computation: a) quantum superposition, b) quantum
entanglement, and c) quantum interference.

In addition to the ability to facilitate the operational requirements
discussed above, a universal digital quantum computer must satisfy the
following criteria:

\bigskip

i)\qquad A distinct set of distributed qubits must be defined that are
individually addressable.

ii) \ \ \ \ \ The qubits must be linkable via a binary interaction.

iii) \ \ \ \ Must be capable of performing two-bit logic operations, i.e.
C-NOT gate.

iv) \ \ \ The speed of operation must satisfy the Preskill criterion, i.e. the
capability must exist for the execution of at \qquad\qquad\ \ \ least 104
distinct operations within the quantum decoherence time.

v) \ \ \ \ \ Arbitrary initial state preparation must be clearly executable.

We present our generic scheme for realization of a quantum computer in the
next section. In section IV we discuss the fulfillment of the above criteria
in our proposed scheme.

\section{III. Realization of a Universal Quantum Computer}

Our proposed scheme is similar to the seminal proposal of Kane[15], but does
not depend upon electronic charge migration for read-in and read-out, nor wave
function displacement to distinguish nuclear spin qubits. Our scheme, that we
feel offers a viable alternative, is represented in fig.2.%

\par\noindent Fig. 2. Schematic representation of distributed qubit array as discussed in
the text. Atoms C and A are embedded in the Si wave guide as shown. The Stark
gates are depicted above atoms C external to the guide, and the bias gates,
located on the top and edge of the guide are also shown.

\bigskip

As depicted in fig. 2 the two-qubit system is constituted by the unpaired
single electron spin of atom C, and the nuclear spin of atom A. The
specification of the atomic constituents is that atom C be composed of a
single loosely bound unpaired electron and isotope of nuclear spin $I_{C}=0$,
whereas, atom A is composed of isotope with $I_{A}=1/2$ and no unpaired
electrons. The configuration of atoms A and C, shown in fig. 2, is embedded in
an optical waveguide composed of pure Si of isotope $I_{SI}=0$. The substrate
also consists of Si with $I_{SI}=0$ to avoid background interference. The
electron spin of atom C is coupled to the nuclear spin of atom A by electronic
wavefunction overlap as shown in fig 2. The coupling is via Fermi contact
interaction[18] given generically for isotropic transferred hyperfine
interaction by
\begin{equation}
A_{s}=\frac{8\pi\beta\hbar\beta_{n}}{3S}\left[  a_{1s}^{2}\left|  \psi
_{1s}\left(  0\right)  \right|  ^{2}+a_{2s}^{2}\left|  \psi_{2s}\left(
0\right)  \right|  ^{2}+2a_{1s}a_{2s}\left|  \psi_{1s}\left(  0\right)
\right|  \left|  \psi_{2s}\left(  0\right)  \right|  \right]
\end{equation}

where $\beta$ is the Bohr magneton, $\beta_{n}$ the nuclear magneton, and the
wave functions $\psi_{ks}(0)$ , k = 1, 2, are the 1s and 2s orbitals of atom A
evaluated at its nucleus, and the a2s=are overlap integrals coupling the
electron wave function of atom C with the 1s and 2s inner shell wave functions
of atom A. Thus, the unpaired electron spin of atom C is coupled with the
nuclear spin of atom A by contact interaction mediated by the inner shell s -
state orbitals of atom A.

The interaction is controlled by laser field induced electronic excitation of
atom C from its ground s - state (no overlap interaction with atom A) to an
electronic excited state with a p -, or d- orbital, with strong overlap and
superhyperfine interaction with one or both of its nearest neighbor atoms A.
Enhanced, as well as suppressed, interaction is controlled by gates mounted
external to the waveguide, that produce a positive or negative bias. Outer
shell, loosely bound, electronic wave functions may be as extensive as 20nm in
a solid of sufficiently high dielectric constant, so nearest neighbor C- A
atom configurations must be within this range. Selective excitation is
provided by a gate that sandwiches atom C used to Stark shift atom C into and
out of resonance with the cw laser field in the waveguide, thus enabling $\pi$
- excitation / deexcitation. The nuclear spin flips are controlled by
microwave induced simultaneous electron - nuclear spin flips mediated by the
transferred hyperfine interaction.

The elementary, time independent, spin Hamiltonian for this system is,
\begin{equation}
H=g\beta H_{0}S_{z}+S.A.I.-g_{n}\beta_{n}H_{0}I_{z}%
\end{equation}

where $g$ and $g_{n}$ are the electron and nuclear gyromagnetic ratios. Here,
we have omitted the weak transverse time dependent part of the Hamiltonian for
simplicity. The first and third terms are the Zeeman terms for the electronic
and nuclear spins, respectively, coupling to the constant magnetic field,
$H_{0}$. The second term expresses the transferred hyperfine coupling between
the electron spin S of atom C and the nuclear spin I of atom A. These are
coupled via the superhyperfine tensor A, that for isotropic homogeneous
interaction is diagonal in the representation where $S_{z}$ and $I_{z}$ are
aligned with $H_{0}$, and is given in magnitude by (6). For simplicity we
assume isotropy in what follows unless stated otherwise. To first order, and
with the condition
\begin{equation}
g\beta>A_{s}>>g_{n}\beta_{n}%
\end{equation}
the eigen energies of (7) with associated quantum numbers, $m_{s}$ and $m_{I}
$ for electron spin and nuclear spin in the representation in which the Zeeman
terms are diagonal are given by,
\begin{equation}
E(m_{s},m_{I})=g\beta H_{0}m_{s}+A_{s}m_{s}m_{I}%
\end{equation}%

\par\noindent Fig. 3. Zeeman and superhyperfine energy level splitting and microwave
frequency transition for simultaneous electron-nuclear spin flips.

\bigskip

The corresponding Zeeman energy level diagram is shown in fig. 3. It is seen
that simultaneous electron-nuclear spin flips can be induced by an applied
microwave field of frequency $\omega$, according to the selection rule $\Delta
m_{F}=0$, where F = S + I is the total spin angular momentum. Thus the
$E_{4}\rightarrow E_{2}$ transition, corresponding to simultaneous electron -
nuclear spin flips, is allowed under the proper selection rules, $\Delta
m_{F}=\pm1,0$, but the $E_{3}\rightarrow E_{1}$ transition is forbidden.
Explicitly, for an isotropic tensor A, i.e. diagonal in this representation,
transitions are induced according to the second term in (7) by the operators
$S^{\pm}I^{\mp}$, consistent with the selection rule for simultaneous electron
- nuclear spin transitions indicated in fig. 3. However, provided the symmetry
is such that A is not diagonal in this representation, or if there is an
element of symmetry breaking, then the superhyperfine interaction described by
the second term in (7) can contain contributions from mixed terms $S_{x}S_{y}%
$, $S_{y}S_{x}$ that can cause the transition $E_{3}\rightarrow E_{4}$. Such
conditions can, in principle, be regulated by local impurities or the natural
local symmetry environment.

With these conditions, simultaneous electron - nuclear spin flips can be
controlled by externally applied microwave pulses mediated by transferred
hyperfine interactions. The interaction can be regulated, i.e. turned on -
off, by laser pulse electronic excitation / deexcitation between the ground s
- state of atom C (no electronic wave function overlap with the nuclear spin
of atom A) and an electronic excited state of nonzero orbital angular momentum
(strong overlap with the nucleus of atom A) as indicated in fig. 4. This
introduces also the option of storing information in the nuclear spin system.
Most important, however, is the single qubit selectivity facilitated by
selective laser pulse excitation using the Stark shift gates discussed
earlier. In the next Section we discuss some useful specific performance
aspects of the model.%

\par\noindent Fig. 4. Zeeman and superhyperfine (shf) energy level splittings and laser and microwave

pulse sequence for shf interaction control and simultaneous electron- nuclear
spin flips.

\bigskip

\section{IV. Quantum Calculational Features}

With reference to figs. 3-4 we identify the calculational basis,
\begin{equation}
\left|  \downarrow\right\rangle _{e}\left|  \uparrow\right\rangle _{n},\left|
\uparrow\right\rangle _{e}\left|  \downarrow\right\rangle _{n},\left|
\downarrow\right\rangle _{e}\left|  \downarrow\right\rangle _{n},\left|
\uparrow\right\rangle _{e}\left|  \uparrow\right\rangle _{n}%
\end{equation}

where $\left|  {}\right\rangle _{e}$ and $\left|  {}\right\rangle _{n}$
represent electron, and nuclear spin states, respectively. The Zeeman split
electron spin manifold depicted in fig. 3 can be driven by a microwave pulse
of frequency $\omega$ into a coherent superposition described by a single spin
rotation unitary transformation, U(t),
\begin{equation}
U(t)\left|  \downarrow\right\rangle _{e}\left|  \uparrow\right\rangle
_{n}=\alpha(t)\left|  \downarrow\right\rangle _{e}\left|  \uparrow
\right\rangle _{n}+\beta\left|  \uparrow\right\rangle _{e}\left|
\downarrow\right\rangle _{n}%
\end{equation}
where $|\alpha|^{2}+|\beta\}^{2}=1.$ The operation (10a), together with
\begin{equation}
U(t)\left|  \downarrow\right\rangle _{e}\left|  \downarrow\right\rangle
_{n}=\alpha(t)\left|  \downarrow\right\rangle _{e}\left|  \downarrow
\right\rangle _{n}+\beta\left|  \uparrow\right\rangle _{e}\left|
\downarrow\right\rangle _{n}%
\end{equation}
form a quantum controlled NOT-gate (C-NOT-gate). The target bit $\left|
{}\right\rangle _{n}$ is flipped contingent upon the state of the control bit
$\left|  {}\right\rangle _{e}$ .The externally applied coherent microwave
pulse induced unitary transformation drives the electron-nuclear spin system
ground and excited states into an entangled pair via the transferred hyperfine
interaction, and the entanglement is manifestly nonlocal. Since two-bit gates
are universal in quantum computing [19], the identification of the set of
qubits and the calculational basis, together with the demonstrated C-NOT gate,
(12), fulfills the requirement for universal quantum computation, i.e., any
algorithm is executable in the system provided the system is scalable.

Worth noting here is that if the target state is initially $\left|
\uparrow\right\rangle _{n}$ , then the unitary transformation is to the
two-bit sub-manifold, $\{\left|  \downarrow\right\rangle _{e}\left|
\uparrow\right\rangle _{n},\left|  \uparrow\right\rangle _{e}\left|
\downarrow\right\rangle _{n}\}$, (11); whereas, given the target state
$\left|  \downarrow\right\rangle _{n}$, the same transformation results in the
transformation to the submanifold, $\{\left|  \downarrow\right\rangle
_{e}\left|  \downarrow\right\rangle _{n},\left|  \uparrow\right\rangle
_{e}\left|  \uparrow\right\rangle _{n}\}$. These alternatives are isomorphic
to the oracle of the Deutsch-Jozsa Promise Algorithm [20], and the initial
nuclear spin target state can be determined by application of that algorithm
in a single unitary operation.

A different calculational basis can be identified, independently, with respect
to the nuclear spins. Selective rf field induced nuclear spin flips can be
induced by laser pulse controlled transferred hyperfine interaction. The
selective transferred hyperfine interaction can be used to induce significant
nuclear spin level shifts. This interesting alternative and option will not be
pursued further here, but will be treated elsewhere. We feel that we have
presented here the simplest approach to quantum computation with respect to
this particular scheme.

So far, we have discussed results for selective coupling between an atom C,
with respect to its electronic component, and a single atom A, with its
nuclear component. Now we focus attention on electronic wave function mediated
coupling of two adjacent nuclear spins, fig. 2. For this case, instead of (7)
we have the Hamiltonian
\begin{equation}
H=g\beta H_{0}S_{z}+A_{k}S.I_{k}+A_{k+1}S.I_{k+1}-g_{nk}\beta H_{0}I_{z}%
^{(k)}-g_{nk+1}\beta_{n}H_{0}I_{z}^{(k+1)}%
\end{equation}
and to first order and within the approximation (8), the associated energy
levels are given by
\begin{equation}
E(m_{s},m_{Ik},m_{Ik+1})=g\beta H_{0}m_{s}+A_{s}m_{s}(m_{I}+m_{Ik+1})
\end{equation}
Here, k labels the location of atom $A_{k}$, and $m_{Ik}$, and $m_{Ik+1}$ are
the spin quantum numbers for nuclei of atoms $A_{k}$ and $A_{k+1}%
$respectively. The corresponding energy level diagram and transitions are
displayed in fig. 5.%

\par\noindent Fig.5 Zeeman and superhyperfine energy level splittings and microwave field
induced transitions for electron spin mediated nuclear spin flips (see fig.
2). Here, the left-hand arrow corresponds to the nuclear spin of the kth atom
of type A whereas the right-hand arrow corresponds to the nuclear spin of the
k+1 atom of type A.

\bigskip

\bigskip

As an example, consider the $\Delta m_{I}=-1$ microwave excitation from the
ground initial state
\begin{equation}
\left|  \psi_{l}\right\rangle =\left|  \downarrow\right\rangle _{e}\left|
\uparrow\uparrow\right\rangle _{n}%
\end{equation}
to the state
\begin{equation}
U_{1}(t)\left|  \downarrow\right\rangle _{e}\left|  \uparrow\uparrow
\right\rangle _{n}=\left|  \downarrow\right\rangle _{e}\left|  \uparrow
\downarrow+\downarrow\uparrow\right\rangle _{n}%
\end{equation}
Corresponding to the transition $E_{6}\longrightarrow E_{2}$ , fig.5. Here,
the left arrow in $\left|  {}\right\rangle _{n}$ always refers to atom $A_{k}$
, and the right arrow refers to atom $A_{k+1}$ nuclear spin. This is followed
by the $\Delta m_{I}=0$ microwave pulse deexcitation,
\begin{equation}
U_{1}(t)\left|  \uparrow\right\rangle _{e}\left|  \uparrow\downarrow
+\downarrow\uparrow\right\rangle _{n}=\left|  \downarrow\right\rangle
_{e}\left|  \uparrow\downarrow+\downarrow\uparrow\right\rangle _{n}%
\end{equation}
Corresponding to the transformation $E_{2}\longrightarrow E_{5}$ .
Subsequently, this is followed by laser pulse induced adiabatic electron
spin-nuclear spin decoupling
\begin{equation}
\left|  \downarrow\right\rangle _{e}\left|  \uparrow\downarrow+\downarrow
\uparrow\right\rangle _{n}\longrightarrow\left|  \uparrow\downarrow
+\downarrow\uparrow\right\rangle _{n}%
\end{equation}%
\begin{equation}
A_{s}\longrightarrow0
\end{equation}
thus storing entangled state information in the nuclear spin system,
\begin{equation}
\left|  \psi_{f}\right\rangle _{n}=\left|  \uparrow\downarrow+\downarrow
\uparrow\right\rangle _{n}%
\end{equation}

Now, suppose the superhyperfine coupling is again induced adiabatically via
laserpulse electronic excitation, except that the overlap is enhanced with
respect to the left- hand atom $A_{n}$, and at the same time suppressed with
regard to the right hand atom, $A_{n+1}$ by means of the outer gate elements,
fig. 3. The electron spin of atom C is now coupled to the nuclear spin of atom
$A_{n}$ only. This is represented by
\begin{align}
\left|  \uparrow\downarrow+\downarrow\uparrow\right\rangle _{n}  &
\longrightarrow\left|  \downarrow\right\rangle _{e}\left|  \uparrow
\downarrow+\downarrow\uparrow\right\rangle _{n}\\
A_{s}  & \neq0
\end{align}
Then, microwave excitation from the initial state (18) gives
\begin{equation}
U_{1}(t)\left|  \downarrow\right\rangle _{e}\left|  \uparrow\downarrow
+\downarrow\uparrow\right\rangle _{n}=\left|  \uparrow\right\rangle
_{e}\left|  \uparrow\downarrow+\downarrow\uparrow\right\rangle _{n}%
\end{equation}
Then, adiabatic laser pulse induced electron-spin, nuclear-spin decoupling
yields the entangled nuclear spin state,
\begin{equation}
\left|  \psi_{f}\right\rangle _{n}=\left|  \uparrow\uparrow+\downarrow
\downarrow\right\rangle _{n}%
\end{equation}
Thus, we have manipulated the system to induce two distinct entangled pairs of
spin states, (20) and (24), and this information can be stored in the nuclear
spin system. In a similar manner we can produce a uniform superposition of
(20) and (24).

\section{V. Universality}

\bigskip

We are now in position to discuss the universality of our scheme in relation
to the criteria of Section II.: i) We have identified a distinct set of qubits
and a computational basis, (10), fig.2; ii)The qubits are distinct,
distributed, and individually addressable, fig. 3; iii) The coupling of
electron and nuclear spins is via transferred hyperfine interaction, (5-6),
fig. 3; iv) We have demonstrated a C- NOT gate, (11,12); v) The speed of
operation is governed by the rate at which an externally applied microwave
field can cause simultaneous electron-nuclear spin flips. This rate is limited
by the strength of the transferred hyperfine interaction, As 10KHz - 100MHz.
The nuclear spin flip relaxation times in Si:31P are measured to be within the
1-10 hour range at low temperatures [21], whereas the direct electron spin
relaxation is found to be on the order of one hour. The electron spin
resonance (ESR) line width for 106 Phosphorus ions / cm3 in Si was observed to
be 1kHz using spin-echo techniques [22], much narrower than the ESR frequency
on the order 10 GHz. Thus, Preskill's criterion [23] is well satisfied for the
configuration of our QC scheme; vi) We demonstrated selective single qubit
addressability in Section IV. Initial state preparation can be established
either globally by the usual techniques of electron-nuclear spin resonance,
NMR, and ESR, or by selective single bit preparation using laser pulse
selective electronic excitation and subsequent microwave pulse initial state
preparation. There are a variety of combinations of techniques that can be
used; however, the selective single qubit preparation is sufficient to
guarantee arbitrary initial state preparation capability; vii) Input
information can be imparted selectively to either the nuclear or electronic
spin systems, or both, by microwave pulse excitation following selective laser
pulse electronic excitation. Output information can be rendered efficiently
via optical fluorescence from electronic excited states. Output information is
therefore imparted to the excited state electronic spin system. The
fluorescence yield conveys all information involving the electron-spin
nuclear-spin system. The specifics of optical fluorescence and detection for
this system will be the subject of another publication [24]; vii) scalability
was clearly identified in the previous Section, (15-24).

\section{\bigskip VI. Summary and Conclusions}

\bigskip

We have demonstrated a scheme for quantum computing and shown that it
satisfies the irreducible set of criteria for universality. Our scheme has
definitive advantages over other schemes involving solid state systems,
notably that of Kane [15], and Yablonovitch [25]. These schemes, each, involve
electron charge transfer and readin / readout requiring single electron charge
measurements. Our method is more robust, depending upon microwave pulse
excitation control for reading and sensitive optical fluorescence and single
photon detection for readout.

Fabrication technology requirements for the scheme proposed here is currently
beyond the state-of-the-art, but this current impasse is expected to be
alleviated within a few years due to the rapid progress in Nanoscience of
materials and fabrication, especially in regard to silicone fabrication
technology. Another concern involves the specific choice of atomic species to
satisfy the conditions required for the qubit design. This requires
simultaneous materials study and optimization analysis. The spectroscopy
techniques required in this scheme provide a particular challenge. The NMR,
ESR, and double resonance requirements correspond to well developed
techniques; however, this combined with integrated laser pulse excitation adds
a third component to constitute a triple resonance spectroscopy. This may lead
to the cultivation of an interesting and useful spin-off as a novel method in spectroscopy.

It is felt that the proposed scheme combines several novel ideas to provide a
realistic multi-purpose method of exploiting quantum parallelism,
entanglement, and interference. We anticipate that the scheme and analysis
expressed here can serve as a template for advancement toward realization of a
practical universal quantum computer.

\bigskip

\section{References}

\bigskip

1. Feynman, R. P., 1982, Int. J. Theor. Phys.21, 467.

2. Deutsch, D., 1985, Proc. Roy. Soc. London, Ser. A 400, 96.

3. Shor, P., 1994, Proc. 35th Annual Symp. On Foundations of Computer Science,
Los Alamitas, CA, IEEE \qquad\qquad\qquad\qquad\ \ \ \ Press, 124.

4. Ekert, A., and Jozsa, R., 1996, Rev. Mod. Phys. 68, 733.

5. Grover, L. K., 1997, Phys. Rev. Lett. 78, 325.

6. Turchette, Q. A., Hood, C. J., Lange, W., Mabuchi, H., and Kimble, H. J.,
1995, Phys. Rev. Lett. 75, 4710.

7. Cirac, J. I., and Zoller, P., 1995, Phys. Rev. Lett. 74, 4091.

8. Monroe, C., Meekhof, D. M., King, B. E., Itono, W. M., ans Wineland, D. J.,
1995, Phys. Rev. Lett. 75, 4714.

9. Gershenfeld, N. A., and Chuang, I. L., 1997, Science 275, 350.

10. Chuang, I. L., Gershenfeld, N. A., and Kubinec, M., 1998, Phys Rev. Lett.
80, 3408.

11. Bowden, C. M., Dowling, J. P., and Hotaling, S. P., 1997, Proc. SPIE 1111
Annual International Symposium on Aerospace / Defence, Sensing, Simulations,
and Control, Orland, FL, April.

12. Bowden, C. M., and Miller, J. E., 1967, Phys. Rev. Lett. 19, 4.

13. Marshall, W., and Stuart, R., 1961, Phys. Rev. 123, 2048.

14. DiVincenzo, D. P., 1998, Nature 393, 113.

15. Kane, B. E., 1998, Nature 393, 133.

16. Cleve, R., Ekert, A., Macchiavello, C., and Mosca, M., 1997, quant-ph/9708016.

17. Scully, M. O., and Zubairy, M. S., 1997, Quantum Optics (Cambridge), Ch.4, 17.

18. Slichter, C. P., 1963, (Harper \& Row), Ch. 7.

19. DiVincenzo, D. P., 1995, Phys. Rev. A 51, 1015.

20. Deutsch, D., and Jozsa, R., 1992, Proc. Roy. Soc. London A 439, 553.

21. Feher, G., and Gere, E. A., 1959, Phys. Rev. 114, 1245.

22. Chiba, M. and Hirai, A., 1972, J. Phys. Soc. Japan 33, 730.

23. Preskill, J., 1998, Royal Soc. London, A 454, 385.

24. Bowden, C. M., Pethel, S. D., and Bloemer, M. J., in preparation.

25. Yablonovitch, E., private communication.
\end{document}